\documentclass[preprint,12pt]{elsarticle}

\usepackage{amssymb}
\usepackage{lineno}
\journal{Chaos, Solitons \& Fractals}

\begin{document}

\begin{frontmatter}

\title{Chameleon attractors in turbulent flows}

\author[inst1]{Tommaso Alberti}
\affiliation[inst1]{organization={INAF-Istituto di Astrofisica e Planetologia Spaziali},
            addressline={via del Fosso del Cavaliere, 100}, 
            city={Rome},
            postcode={00133}, 
            country={Italy}}

\author[inst2]{Francois Daviaud}
\affiliation[inst2]{organization={CEA, IRAMIS, SPEC, CNRS URA 2464, SPHYNX},
            city={Gif-sur-Yvette},
            postcode={91191}, 
            country={France}}

\author[inst3,inst4]{Reik V. Donner}
\affiliation[inst3]{organization={Department of Water, Environment, Construction and Safety, Magdeburg–Stendal University of Applied Sciences},
            addressline={Breitscheidstraße, 2}, 
            city={Magdeburg},
            postcode={39114}, 
            country={Germany}}
\affiliation[inst4]{organization={Research Department I – Earth System Analysis, Potsdam Institute for Climate Impact Research (PIK) – Member of the Leibniz Association},
            addressline={Telegrafenberg, A31}, 
            city={Potsdam},
            postcode={14473}, 
            country={Germany}}

\author[inst5]{Berengere Dubrulle}
\affiliation[inst5]{organization={SPEC, CEA, CNRS, Universite Paris-Saclay},
            city={Gif-sur-Yvette},
            postcode={91191}, 
            country={France}}

\author[inst6,inst7,inst8]{Davide Faranda}
\affiliation[inst6]{organization={Laboratoire des Sciences du Climat et de l'Environnement, CEA Saclay l'Orme des Merisiers, UMR 8212 CEA-CNRS-UVSQ, Universit Paris-Saclay \& IPSL},
            city={Gif-sur-Yvette},
            postcode={91191}, 
            country={France}}
\affiliation[inst7]{organization={London Mathematical Laboratory},
            addressline={Margravine Gardens, 8}, 
            city={London},
            postcode={W6 8RH}, 
            country={UK}}
\affiliation[inst8]{organization={LMD/IPSL, Ecole Normale Superieure, PSL research University},
            city={Paris},
            postcode={75005}, 
            country={France}}

\author[inst9,inst10]{Valerio Lucarini}
\affiliation[inst9]{organization={Department of Mathematics and Statistics, University of Reading},
            city={Reading},
            country={UK}}
\affiliation[inst10]{organization={Centre for the Mathematics of Planet Earth, University of Reading},
            city={Reading},
            country={UK}}

\begin{abstract}
Turbulent flows present rich dynamics originating from non-trivial energy fluxes across scales, non-stationary forcings and geometrical constraints. This complexity manifests in non-hyperbolic chaos, randomness, state-dependent persistence and unpredictability. All these features have prevented a full characterization of the underlying turbulent (stochastic) attractor, which will be the key object to unpin this complexity. 
Here we propose a novel formalism to trace the evolution of the structural characteristics of phase-space trajectories across scales, providing a full characterization of the attractor. We demonstrate that the properties of the dynamically invariant objects depend on the scale we are focusing on. In the case of laboratory experiments on fluids we observe the emergence of an intrinsic timescale, solely determined by nonlinear interactions, controlling the geometrical and topological properties of phase-space trajectories. Given the changing nature of such attractors in time and scales we term them {\em chameleon attractors}.
\end{abstract}


\begin{highlights}
\item We introduce appropriate tools to investigate the local in time and space properties of attractors
\item We show that a turbulent attractor is sensitive to the emergence of an intrinsic timescale solely determined by nonlinear interactions
\item The attractor adapts its geometric and statistical properties dynamically in time with respect to the intrinsic timescale, we call such attractor a "chameleon" attractor
\end{highlights}


\end{frontmatter}


\section{Introduction}
The dynamics of turbulent flows is usually described via macroscopic dynamical laws, the Navier-Stokes equations (NSE), derived from the mesoscopic Boltzmann equation \cite{Esposito1999}. They accurately describe the dynamics of averaged quantities over spatial scales much larger than the mean free path length of fluid molecules \cite{Foias01}. Nevertheless, when considering large-scale turbulent flows even the most powerful computer on Earth fails to accurately describe their behavior at all relevant scales. Hence, there is a need to develop accurate yet efficient parameterizations for describing the impact of the unresolved scales of motions on those of interest \cite{Berner2017,Ghil2020}.

One way to approach the description of fluid flows containing a large number of scales is via statistical analysis as in the multifractal formalism developed by Parisi and Frisch \cite{Parisi85}. It is based on characterising the high-order statistics of the velocity field increments, thought to be representative of fluctuations at different scales, via a set of {\em scaling exponents} \cite{Benzi84}. Contrasting the usual idea, coming from critical phenomena, that only a countable set of scaling exponents are relevant for a complete characterization of the statistical features of fluid flows \cite{Crisanti93,Ellis99}, Parisi and Frisch \cite{Parisi85} introduced an infinite hierarchy of exponents, each belonging to a given fractal set. These exponents account for all possible (infinite) rescaling symmetries of the NSE, describing the existence of singularities in the energy cascade mechanism in turbulent flows \cite{Benzi84,Dubrulle19}. Since the development of the multifractal theory experimental measurements of the velocity field in fluids have proved to be compatible with this picture \cite{Muzy91,Benzi91,Biferale04,Boffetta08,Benzi08,Arneodo08}. However, this approach only provides global information on the scale-dependent properties of fluids via the probability of occurrence of a given scaling exponent. Moreover, a direct computation of the multifractal spectrum from the NSE is not possible~\cite{Lanotte15}, although it would be helpful to explore the local statistics of velocity field fluctuations \cite{Dubrulle19}. 

A complementary approach to the high-order statistics is provided in the framework of dissipative chaotic dynamical systems \cite{Lorenz63}, exploiting the fact that the concepts of turbulence and chaos are closely connected \cite{Ruelle71}. Indeed, three-dimensional viscous fluids, as described via the NSE, conform to this class of systems, being characterized by strange attractors, i.e., phase-space states toward which the system evolves for a wide range of initial conditions resulting from a series of bifurcations~\cite{Ruelle71}. However, the search for an attractor underlying turbulent flows has only proved partially successful so far \cite{Takens81,Miles84,Crutchfield88,Bohr05}. Indeed, several studies have suggested that the observed dynamical processes can be associated with the existence of non-hyperbolic strange and possibly stochastic attractors having a dimensionality much lower than the number of degrees of freedom of the system \cite{lucarini2016extremes,LucariniGritsun2020}. Non-hyperbolicity manifests itself with the fact that the attractor is heterogeneous in terms of its local properties of persistence and predictability \cite{LucariniGritsun2020,Faranda17}. When considering numerical models, this has important implications also in terms of error dynamics and efficiency of data assimilation \cite{Vannitsem2016}. 

In this work, we use a laboratory experiment under high Reynolds number turbulent conditions to explore the active number of degrees of freedom at different scales. For this purpose we use a time-dependent parameter providing information on the symmetries of the turbulent steady state, thus allowing us to reconstruct the underlying attractor. By combining a decomposition method, to detect scale-dependent components, with concepts from extreme value theory (EVT), to sample local properties of attractors, we trace the evolution of the geometrical and topological properties of these invariant objects across scales for a symmetric and an asymmetric turbulent state. While the former is characterized by a scale-invariant attractor, the latter is features an attractor that is scale- and time-dependent, being sensitive to the emergence of an intrinsic timescale solely determined by nonlinear interactions. Furthermore, we also demonstrate that the symmetric turbulent steady state is characterized by a simple phase-space topology and geometry, resembling that of a noisy fixed point. Conversely, the asymmetric turbulent steady state displays scale-dependent non-hyperbolic features, moving from a noisy fixed point like structure at small scales (random attractor) towards a two-lobe chaotic attractor at large scales. Thus, because the attractor adapts its geometric and statistical properties dynamically in time with respect to the intrinsic timescale, we call such attractor a {\em chameleon attractor}.

\section{Data}
Our data originate from a turbulent von Karman flow, obtained by stirring rapidly water in a vertical cylinder of length $L = 180$~mm and radius $R = 100$~mm. As a result of the forcing turbulence develops that produces a back-reaction onto the two stirring counter-rotating impellers measured through two torque-meters located along their common axis. The resulting torques $C_1(t)$ and $C_2(t)$ can be seen as large-scale quantities reflecting the complex behaviour of the fluid \cite{SaintMichel13}. Similarly, the instantaneous rotation frequencies $f_1(t)$ and $f_2(t)$ of the two impellers provide a global measure of the large-scale circulation that develops under the action of the angular momentum flux \cite{Thalabard14,SaintMichel14}. Although $f_1(t)$ and $f_2(t)$ provide a 1D (time-only) projection of the full 4D (space-time) dynamics of the turbulent flow, they preserve intrinsic properties of the full turbulent system such as intermittency, bi-stability and, for special forcing conditions, a stochastic attractor. Such a situation is observed when $C_1$ and $C_2$ are constant \cite{Faranda17}; as a result, the two frequencies $f_1(t)$ and $f_2(t)$ fluctuate in time, with a typical mean frequency of $f_0 \sim 7$ Hz \cite{Faranda17}. The corresponding turbulent flow is then characterized by a Reynolds number $Re = 2\pi R^2 f_0 \nu^{-1} \sim 3 \times 10^5$, significantly exceeding the estimated critical Reynolds number for turbulence onset, $Re_T \approx 3500$. However, the time fluctuations of $f_1(t)$ and $f_2(t)$ follow an organized pattern determined by a control parameter $\gamma = \langle (C_1(t)-C_2(t))/(C_1(t)+C_2(t)) \rangle$ and traced by an order parameter $\Theta(t) = (f_1(t)-f_2(t))/(f_1(t)+f_2(t))$. When $\gamma = 0$ the turbulent state is statistically symmetric and $\Theta(t)$ fluctuates around zero \cite{SaintMichel13,SaintMichel14}. For $\gamma \ne 0$, the symmetry is broken and $\Theta(t)$ presents large-scale departures from zero. 
\begin{figure}[h]
  \centerline{\includegraphics[width=\textwidth]{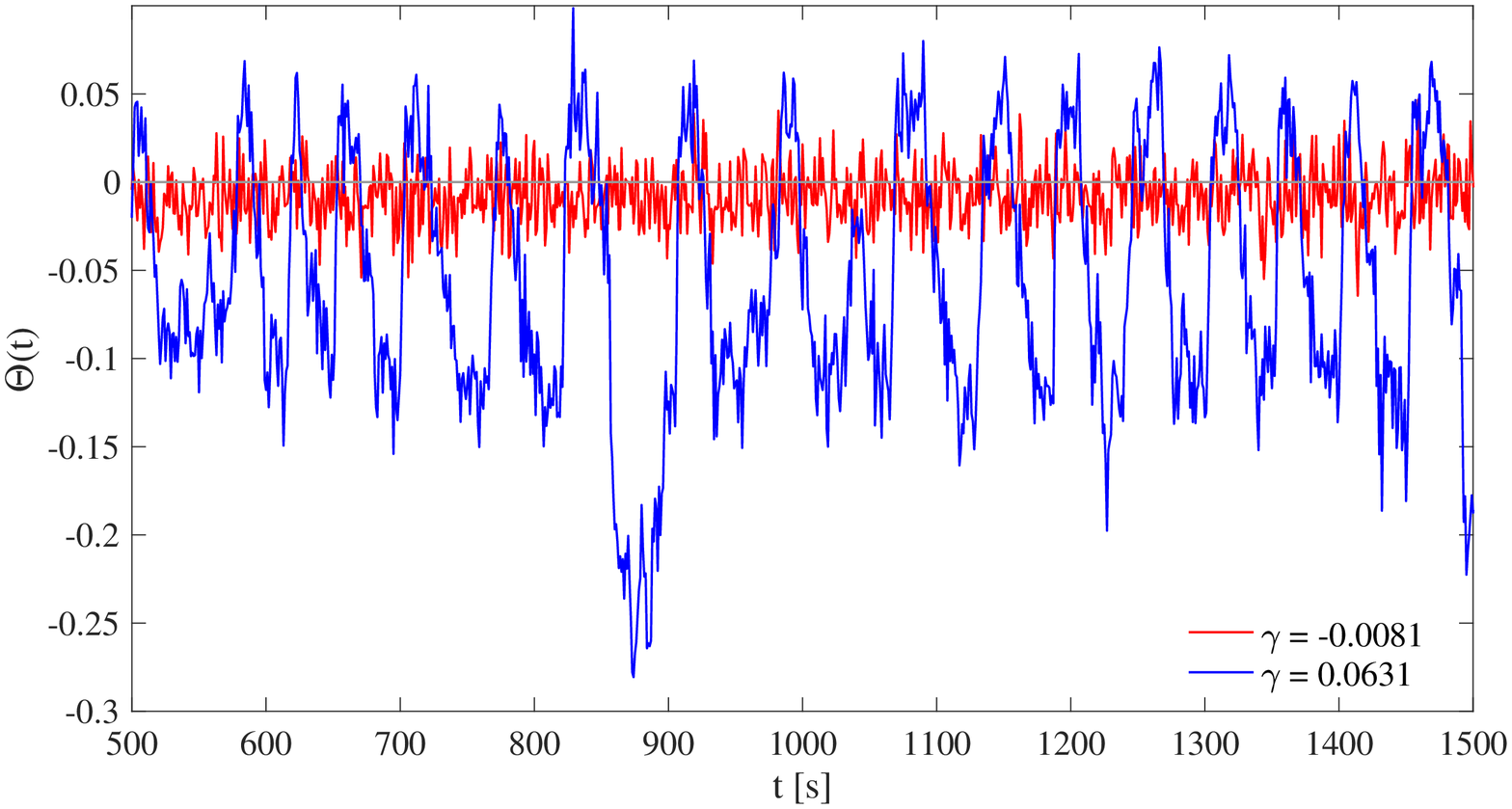}}
  \caption{The temporal behavior of a sample of $\Theta(t)$ for $\gamma = -0.0081$ (red line) and $\gamma = 0.0631$ (blue line). The horizontal gray line refers to $\Theta(t)=0$.}
\label{fig1}
\end{figure}

\begin{figure}[h]
  \centerline{\includegraphics[width=\textwidth]{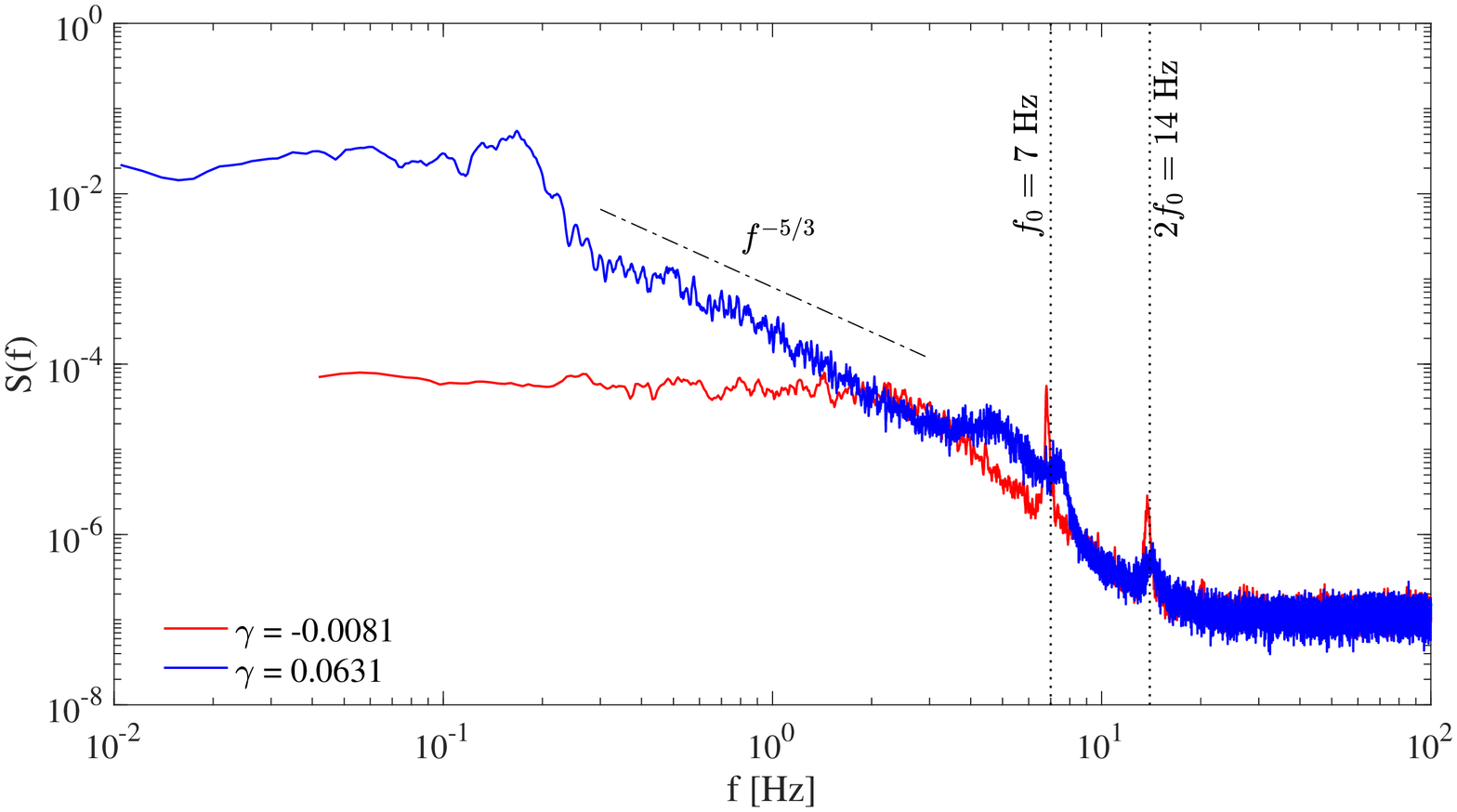}}
  \caption{Power spectral density versus frequency for the two values $\gamma \in \{-0.0081, 0.0631\}$ as reported by red and blue lines, respectively. The vertical dotted lines refer to the typical mean propeller frequency $f_0 = 7$ Hz and its harmonic \cite{Faranda17}.}
\label{fig2}
\end{figure}

Figure~\ref{fig1} reports the time behavior of a sample of $\Theta(t)$ for two selected values of $\gamma$ used in this study,$\gamma = -0.0081$ (symmetric) and $\gamma = 0.0631$ (asymmetric). As expected, we note fluctuations around zero for the symmetric case (i.e., $\gamma = -0.0081$), while large-scale transitions featuring intermittent bursts are found for the asymmetric case (i.e., $\gamma = 0.0631$). The difference between the two values of $\gamma$ can be also highlighted by looking at the spectral properties of $\Theta(t)$ as depicted by the variations of the power spectral density (PSD) across frequencies as reported in Figure \ref{fig2}. 

When $\gamma \sim 0$ the time series resembles that of an uncorrelated white noise, typically characterized by a flat spectrum over a wide range of scales, while for $\gamma > 0$ a turbulent spectrum emerges. Furthermore, the characteristic frequency $f_0 \sim 7$~Hz associated with the average impeller rotation frequency is also recognizable in the spectrum, together with its harmonics at $2 f_0$ \cite{Faranda17}. Then, the spectrum saturates to that of a white noise for $f > 20$ Hz. To account for this behavior, in the following we apply a low-pass filtering procedure with a cut-off frequency $f_{cut} \sim 20$ Hz to our time series to reduce high-frequency fluctuations, also saving computational time in our subsequent calculations \cite{Faranda17}.

\section{Methods}

\subsection{Attractor reconstruction}
\begin{figure}[h]
  \centerline{\includegraphics[width=\textwidth]{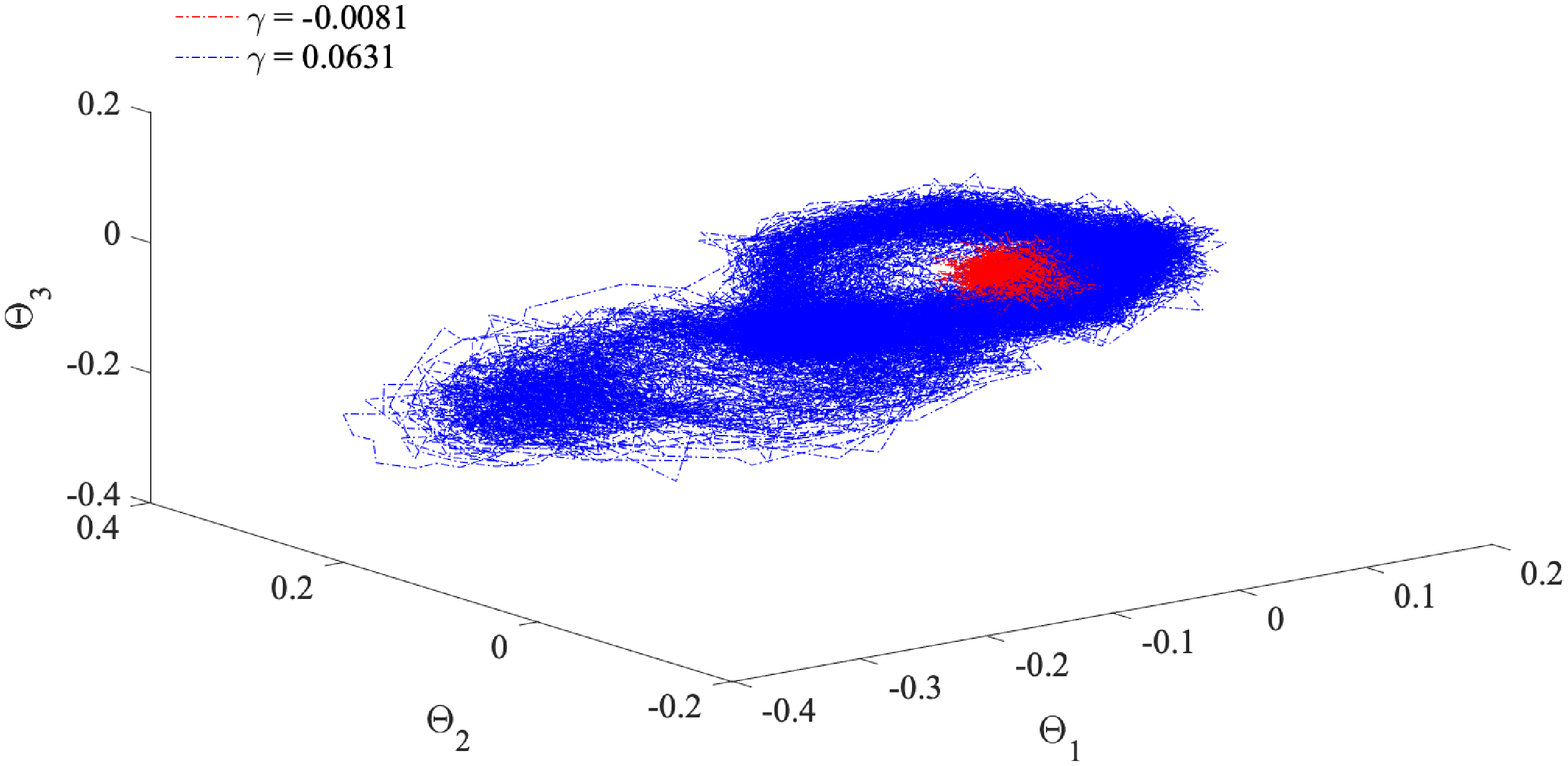}}
  \caption{3-D reconstruction of the full attractor of the system for $\gamma = -0.0081$ (red) and $\gamma = 0.0631$ (blue).}
\label{fig3}
\end{figure}

As shown in Faranda et al. \cite{Faranda17} the dynamical behavior of $\Theta(t)$ can be globally described by a stochastic strange attractor whose geometry depends on $\gamma$. Thus, as a first step of our procedure we reconstruct the global attractor via Takens' embedding method~\cite{Takens81}. This means to translate our univariate representation of the system in terms of the time series $\Theta(t)$ into an $m-$dimensional manifold $\mathcal{M}$ via the following diffeomorphism
\begin{equation}
    \Theta(t) \to \Theta_{m, \Delta}(t) = [\Theta(t), \Theta(t-\Delta), \Theta(t-2\Delta), \ldots, \Theta(t-(m-1)\Delta]^\dagger
\end{equation}
where $\dagger$ indicates the transposition operator. The two parameters, i.e., the embedding dimension $m$ and the time delay $\Delta$, are selected according to standard criteria based on the false nearest neighbor method, suggesting $m = 3$, and the time lag at which the auto-correlation function reduces to 0.5, giving us $\Delta = 20$ time steps \cite[see, e.g.,][]{Faranda17}. In this way, we move from a univariate time series $\Theta(t)$ to a 3-D multivariate signal $\Theta_\mu(t) = [\Theta_1(t), \Theta_2(t), \Theta_3(t)]$. 
The 3-D phase-space for both values of $\gamma$ is reported in Figure \ref{fig3}.

A clear difference emerges between the two reconstructed attractors: while the symmetric case ($\gamma \sim 0$) is characterized by a noisy fixed point like structure, the asymmetric case is clearly characterized by a two-lobe attractor, like that observed for many dissipative chaotic systems \cite{Lorenz63}. However, there is an additional complexity hidden in this global attractor, that reflects the scale dependent properties of turbulence, as we demonstrate in the following.

\subsection{Multivariate Empirical Mode Decomposition (MEMD)}
To uncover the scale dependence, we first decompose the data into intrinsic modes by using the multivariate empirical mode decomposition \cite[MEMD, ][]{Rehman10} that is the multivariate extension of the standard empirical mode decomposition (EMD) \cite{Huang98}. It is an algorithmic procedure directly working in the data domain to detect embedded patterns into multivariate signals $\Theta_\mu(t)$ in the form of so-called Multivariate Intrinsic Mode Functions (MIMFs) \citep{Rehman10}. These patterns are derived through the {\em sifting process} \cite{Huang98}, slightly modified to implement an appropriate cubic spline procedure for multivariate signals \cite{Rehman10}. It consists of the following steps:
\begin{enumerate}
    \item identify local extremes, i.e., points where the $\mu$-variate derivative of $\Theta_\mu(t)$ is zero;
    \item use cubic spline interpolation over these points to derive the upper (maxima) and the lower (minima) envelopes $\mathbf{U}_\mu(t)$ and $\mathbf{L}_\mu(t)$, respectively;
    \item derive the mean envelope $\mathbf{M}_\mu(t) = \frac{\mathbf{U}_\mu(t) + \mathbf{L}_\mu(t)}{2}$ and evaluate the detail $\mathbf{H}_\mu(t) = \Theta_\mu(t) - \mathbf{M}_\mu(t)$.
\end{enumerate}
These steps are iterated until the detail $\mathbf{H}_\mu(t)$ has the same number of extrema and zeros (or having them differing at most by one) and a zero-average mean envelope $\mathbf{M}_\mu(t)$. This means that $\mathbf{H}_\mu(t)$ can be classified as the first Multivariate Intrinsic Mode Function $\mathbf{C}_{\mu, 1}(t)$ (also called multivariate empirical mode) \cite{Huang98,Rehman10}. Then, the algorithmic procedure is repeated over the first residue $\mathbf{R}_{\mu, 1}(t) = \Theta_\mu(t) - \mathbf{C}_{\mu, 1}(t)$ until no more MIMFs $\mathbf{C}_{\mu, k}(t)$ can be filtered out from the data, i.e., the final residue $\mathbf{R}_\mu(t)$ is a $\mu-$variate non-oscillating (monotonic) trend \cite{Rehman10}. Hence, we can write
\begin{equation}
    \Theta_\mu(t) = \sum_{k=1}^N {\bf C}_{\mu, k}(t) + {\bf R}_\mu(t). 
    \label{eq:memd}
\end{equation}
Each ${\bf C}_{\mu, k}(t)$ is a multivariate pattern representative of a peculiar dynamical feature that evolves on a typical multivariate mean timescale $\tau_k$ defined as \cite{Rehman10}
\begin{equation}
\tau_k = \frac{1}{N_p \, \Delta t} \int_{0}^{N_p \, \Delta t} t' \langle \mathbf{C}_{\mu, k}(t') \rangle_{\mu} dt',  
\label{eq:tau}
\end{equation}
where $N_p$ is the number of data points, $\Delta t$ is the time resolution, and $\langle \cdots \rangle$ stands for ensemble average over the $\mu$-dimensional space. MIMFs are by construction ordered in terms of decreasing frequency \cite{Rehman10,Huang98}. Although an {\em a priori} decomposition basis is not fixed, the derived basis, i.e., the set of $\{{\bf C}_{\mu, k}(t)\}$, is a formal mathematical basis, that is, the MIMFs are empirically and locally orthogonal with respect to each other \cite{Rehman10}. Thus, partial sums of Eq.~(\ref{eq:memd}) can be exploited to provide additional information over specific ranges of scales, making the multivariate signal $\Theta_\mu(t)$ interpreted as a superposition of scale-dependent fluctuations \cite{Alberti20}. This property is used in the following to diagnose the dynamical properties of the instantaneous (in time) and local (in phase-space) states.

\subsection{Dynamical system metrics}

The dynamical properties of the $\mu-$variate systems can be investigated by means of two dynamical systems metrics \citep{Lucarini12}, the instantaneous dimension ($d$) and the inverse persistence ($\theta$), based on extreme value theory (EVT). The former is a measure of the active number of degrees of freedom, while the latter is a measure of the short-term stability of the phase-space trajectory associated with the extremal index of the generalized extreme value (GEV) distribution of recurrence distances \citep{Moloney19}. These instantaneous metrics are obtained by sampling the recurrences (i.e., close encounters) of some reference state $\zeta$ and observing that they are distributed according to EVT \cite{lucarini2016extremes,Lucarini12,Lucarini14}.

Formally, let $x(\zeta)$ be the trajectory of the system and let $\zeta^\ast$ be an arbitrary reference state in the phase-space. Let further $g(x(\zeta), \zeta^\ast) = -\log \left[ dist(x(\zeta), \zeta) \right]$ be the logarithmic return, where $dist(x(\zeta), \zeta^\ast)$ is the Euclidean distance between $x(\zeta)$ and $\zeta^\ast$. If we define exceedances as $X(\zeta^\ast) = g(x(\zeta), \zeta^\ast) - s(q, \zeta^\ast)$, with $s(q, \zeta^\ast)$ being an upper threshold corresponding to the $q$--th empirical quantile of $g(x(\zeta), \zeta^\ast)$, the Freitas-Freitas-Todd theorem modified by Lucarini et al. (2014) \cite{Lucarini14} states that the cumulative distribution $F(X, \zeta^\ast)$ of returning to a sphere of radius $r$ around $\zeta^\ast$ converges to the exponential member of the generalized Pareto family
\begin{equation}
    F(X, \zeta^\ast) \simeq \exp \left[ -\theta(\zeta^\ast) \frac{X(\zeta^\ast)}{d^{-1}(\zeta^\ast)} \right],
\end{equation}
where $0 \le d < \infty$ is the local dimension and $0 \le \theta \le 1$ is the inverse persistence of the state $\zeta^\ast$. Since each point of the trajectory of the phase-space corresponds to a time instant of our embedded time series by means of $d$ and $\theta$ we have a time-dependent view of the properties of our system. However, this only provides information on the full structure of the attractor, without revealing additional features that can be related to processes and mechanisms operating at different scales. For this reason, following Alberti et al. (2020) \cite{Alberti20} we firstly use the MIMFs to reconstruct the dynamics at different ranges of frequencies by exploiting partial sums of Eq.~(\ref{eq:memd})
\begin{equation}
    {\bf \Theta}^f_\mu(t) = \sum_{k | f_k = 1/\tau_k > f^\ast} {\bf C}_{\mu, k}(t),
    \label{eq:Btau}
\end{equation}
giving us a description of the dynamical features at frequencies larger than $f^\ast$. Starting from the largest frequency (i.e., $k=1$) and adding lower and lower ones ($k = 2, 3, ..., N)$ we can introduce for each frequency $f$ a scale-dependent instantaneous dimension $D(t, f)$ and inverse persistence $\theta(t, f)$ by diagnosing the dynamical properties of ${\bf \Theta}^f_\mu(t)$. 

\section{Results}
\begin{figure}
  \centerline{\includegraphics[scale=0.6]{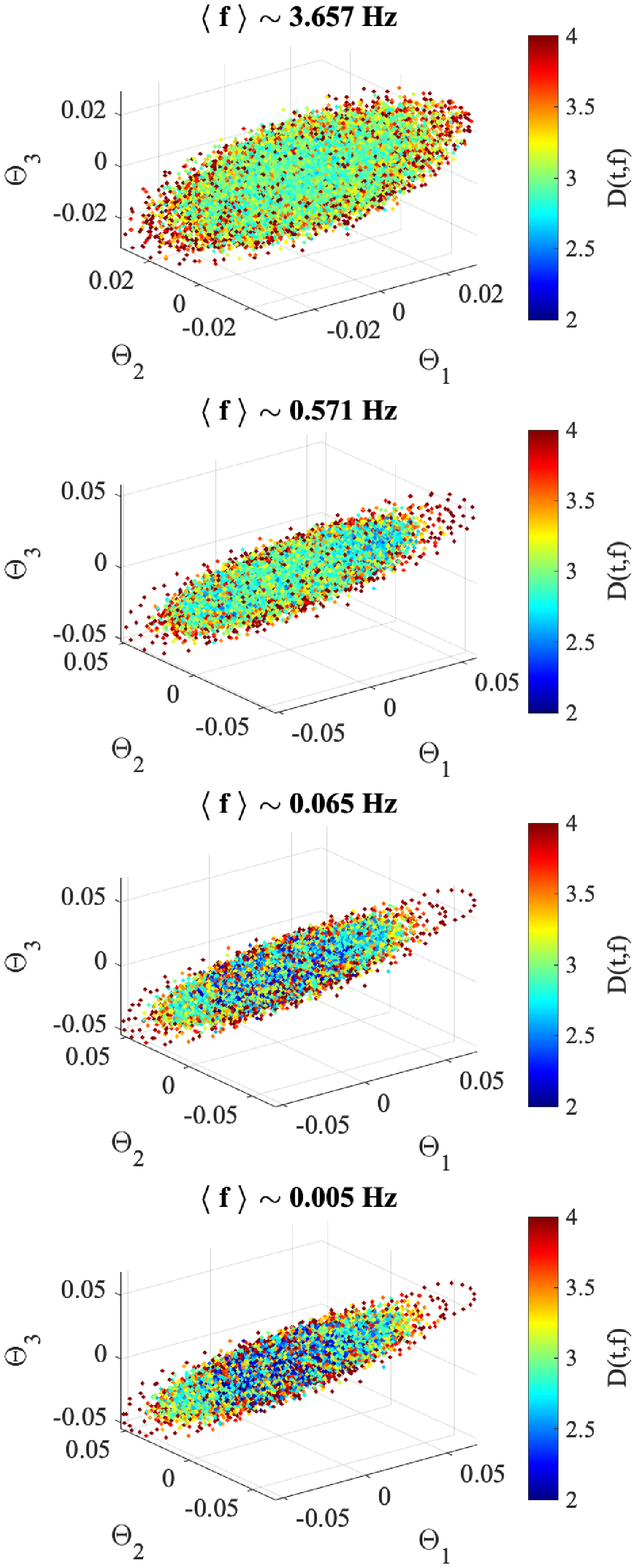}\includegraphics[scale=0.6]{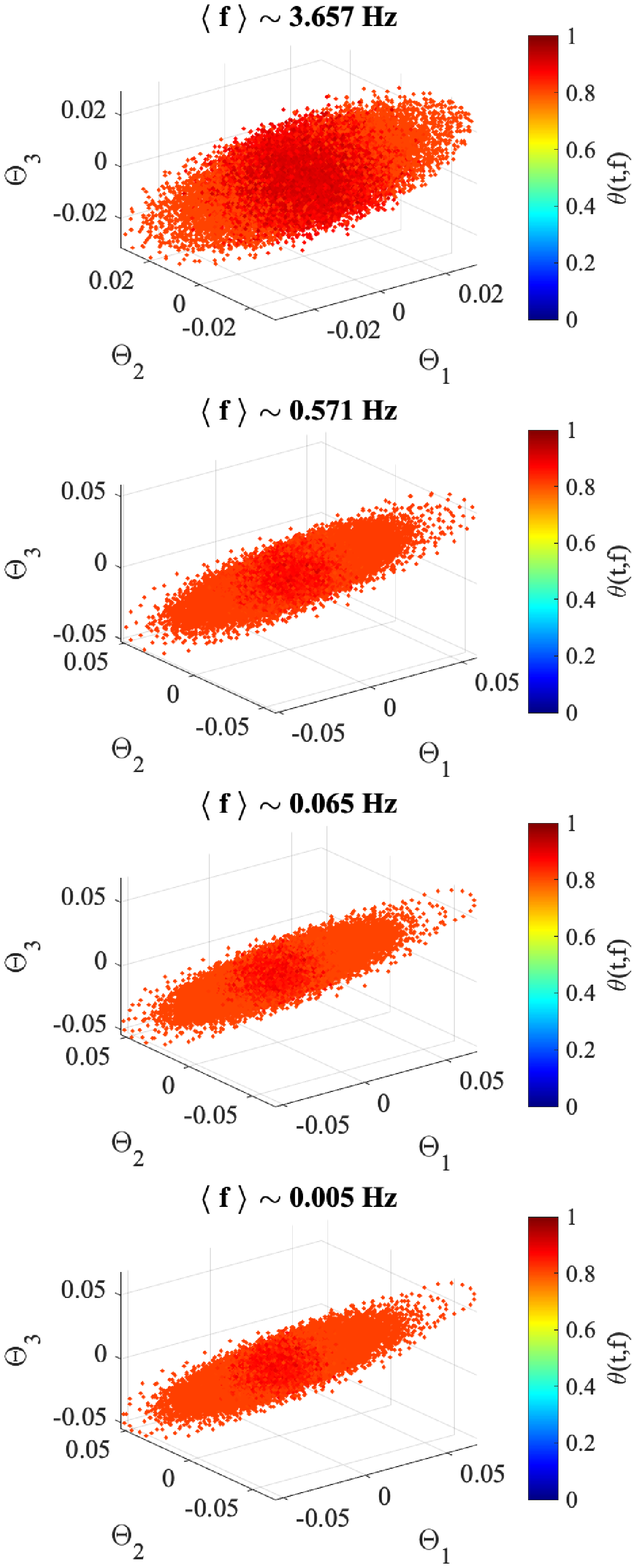}}
  \caption{The 3-D attractor of the system at different frequencies colored by, respectively, the scale-dependent instantaneous dimension $D(t, f)$ (left panels) and the scale-dependent inverse persistence $\theta(t, f)$ for the symmetric case $\gamma = -0.0081$. Moving from top to bottom we consider high to low cutoff frequencies. The color bar for the left panels is saturated between 2 and 4 for visual purposes.}
\label{fig4}
\end{figure}

\begin{figure}
  \centerline{\includegraphics[scale=0.6]{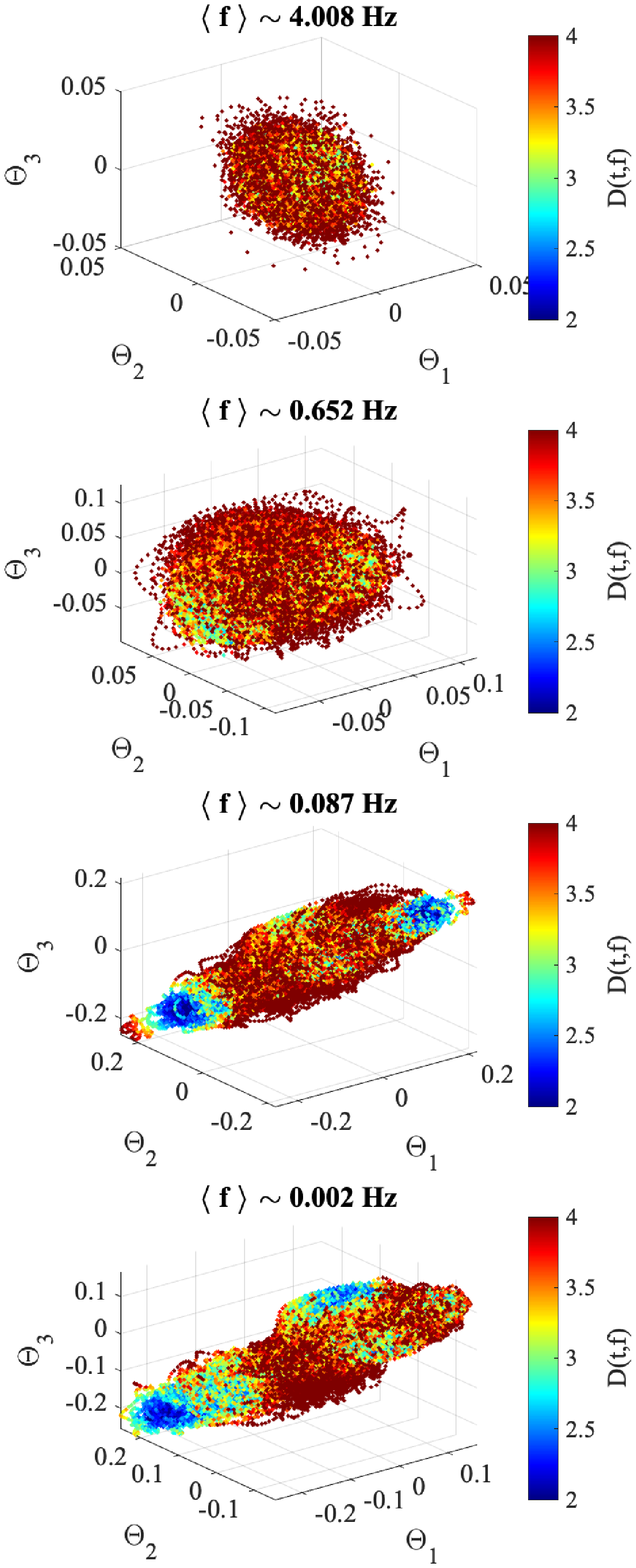}\includegraphics[scale=0.6]{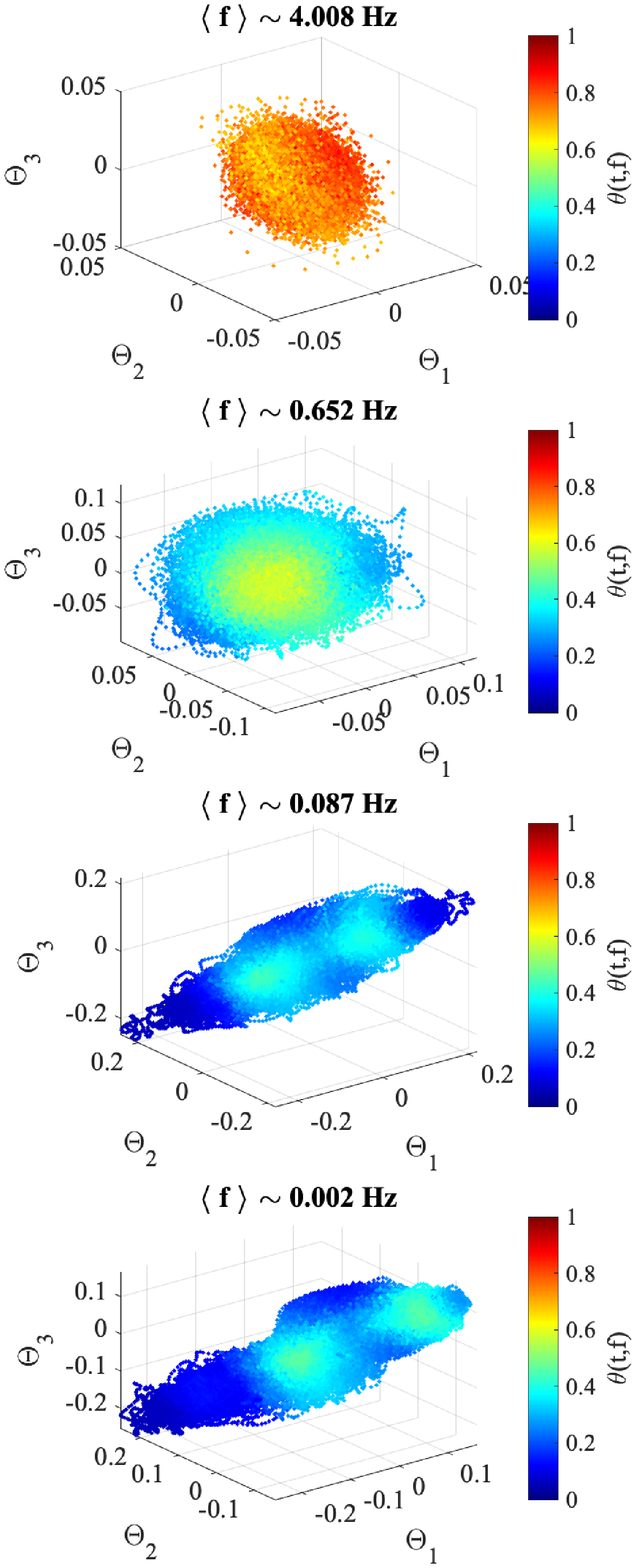}}
  \caption{Same as Figure \ref{fig4} but for the case $\gamma = 0.0631$.}
\label{fig5}
\end{figure}

Figures~\ref{fig4} and \ref{fig5} show the instantaneous dynamical system metrics for (a) the symmetric case with $\gamma = -0.0081$ and (b) a case with full symmetry breaking at $\gamma = 0.0631$, respectively. 

In the symmetric case (Figure \ref{fig4}), we observe that the geometrical properties of the attractor in phase-space are completely invariant with respect to frequency, suggesting that the properties of the system do not depend on the scale. The scale-dependent instantaneous dimension $D(t, f)$ slightly depends on frequency, although the most probable and the average value $\langle D(t, f) \rangle \approx 3$ are about the same  for all frequencies. The scale-dependent inverse persistence $\theta(t, f)$ is mostly characterized by values larger than 0.8, with an average value $\langle \theta(t, f) \rangle \approx 1$ for all frequencies as expected for an unstructured stochastic system \cite{Faranda17}. By contrast, in the non-symmetric case (Figure \ref{fig5}), this scale-invariance is broken. As a result, we observe sudden bursts of scale-dependent instantaneous dimensions $D(t, f) \ge 6$, temporally localized differently at different frequencies. The scale-dependent inverse persistence $\theta(t, f)$ displays also markedly different behaviour across frequencies with respect to the symmetric case ($\gamma \sim 0$), with values close to one at high frequencies and less than 0.2 at lower ones. Furthermore, we clearly observe a transition towards a two-lobe attractor at frequencies less than 0.2 Hz, thus matching the expected phase-space geometry of the full attractor as well as the break in the scaling observed in the PSD (see Figure \ref{fig2}). Thus, while in the symmetric case, $\gamma = -0.0081$, we find that the $D(t, f)$ and $\theta(t, f)$ are homogeneously distributed across the attractor, implying that its topology is very simple and compatible with a noisy fixed point, the non-symmetric case attractor displays scale-dependent features with a heterogeneous spatial distribution of the two metrics.

\begin{figure}
  \centerline{\includegraphics[width=\textwidth]{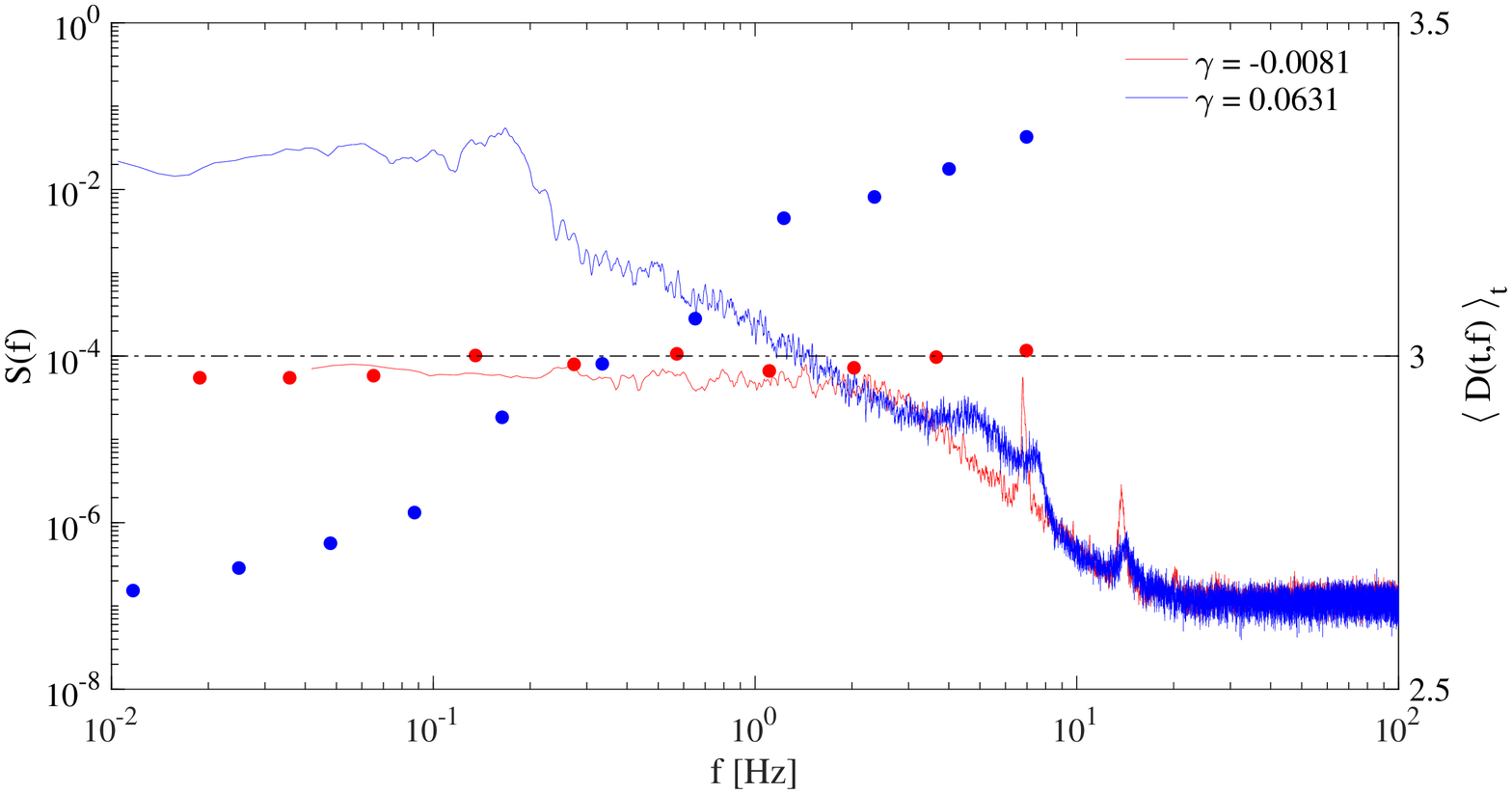}}
  \centerline{\includegraphics[width=\textwidth]{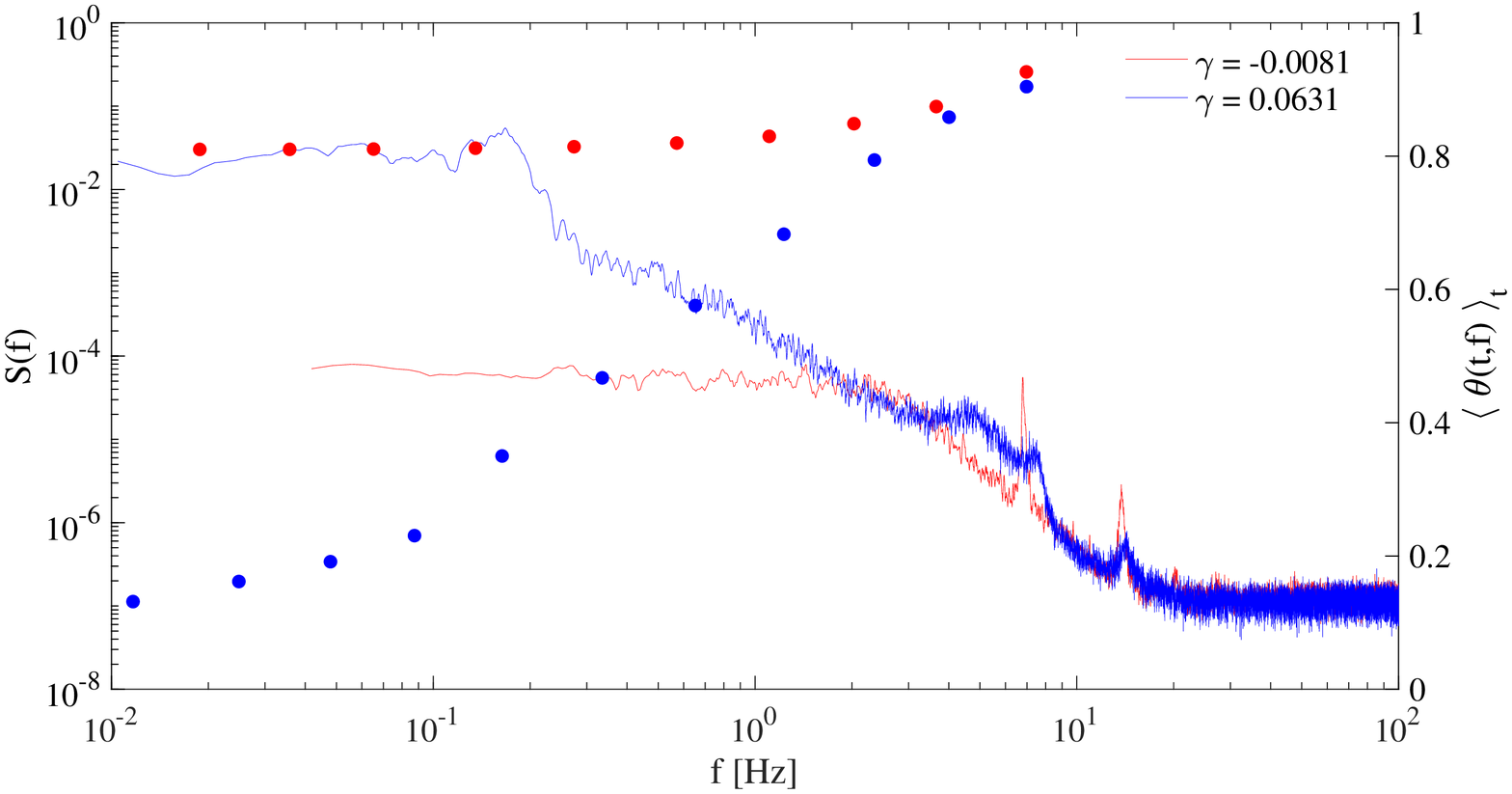}}
  \caption{The behavior of the average dimension $\langle D(t, f) \rangle$ (upper panel, filled circles) and persistence $\langle \theta(t, f) \rangle$ (lower panel, filled circles) in comparison with the PSD (reported as solid lines). Red symbols/lines refer to $\gamma = -0.0081$, while blue symbols/lines refer to $\gamma = 0.0631$.}
\label{fig6}
\end{figure}

To further highlight the scale-dependent features, we show in Figure \ref{fig6} the behavior of the average dimension $\langle D(t, f) \rangle$ and persistence $\langle \theta(t, f) \rangle$ in comparison with the PSD. 
We clearly observe that, on average, the symmetric case presents a scale-invariant behavior of the two metrics, with values close to those expected for a stochastic system (i.e., $\langle D(t, f) \rangle = 3$ and $\langle \theta(t, f) \rangle = 1$). Conversely, a scale-dependent behavior is observed in the non-symmetric case, with a transition between $\langle D(t, f) \rangle < 3$ and $\langle D(t, f) \rangle > 3$ occurring around the low-frequency break observed in the PSD. This also corresponds to changes from  $\langle \theta(t, f) \rangle < 0.5$ to $\langle \theta(t, f) \rangle \to 1$. Thus, our findings clearly suggest that we are faced with a scale-dependent modification of the geometrical and topological properties of the underlying attractor, depending on the emergence of an intrinsic timescale solely determined by nonlinear interactions. This means that we observe a time behavior mirroring the well-known scaling behavior of a 3-D turbulent flow. At scales larger than the injection scale, the energy transfer is small, and the individual scales are in quasi-equilibrium; for scales smaller than the injection scale, the mean energy transfer is positive, and there is an out-of equilibrium energy cascade towards smaller scales, following a Kolmogorov spectrum with intermittency corrections \cite{Kolmogorov41}. In the present case, the low frequencies are associated to low-dimensional dynamics, showing that the statistical equilibrium at large scales is driven by a few degrees of freedom, generating a well defined low-dimensional attractor. On the other hand, the dynamics at scales smaller than the injection scale effectively plays the role of noise, which restores the broken symmetry and provides the "statistical temperature" for large scales, or the stochasticity of the attractor \cite{Thalabard14}. Finally, since our reconstructed 3D phase-space defined via $\Theta_\mu(t))$ is just a projection of a higher-dimensional attractor where other degrees of freedom are lump in stochastic terms (i.e., at small scales), it is not surprising that we find dimensions larger than 3. As shown in Faranda et al. (2017) \cite{Faranda17}, this points towards the existence of an unstable fixed point associated with abrupt changes and hints at the existence of an underlying stochastic attractor. Our scale-dependent results also suggest that, although the flow dynamics involves a wide range of scales, some of them can be described by stochastic theory \cite{Alberti21}. 

\section{Conclusions}

In this paper we have introduced a new multiscale analysis tool to deal with the investigation of time and scale-dependent properties of a simple system derived from a turbulent flow. By exploiting two different cases with different properties in terms of symmetry, we find evidence of a scale-invariant nature of the geometrical properties of the phase-space for the symmetric case. Conversely, in the non-symmetric case the scale-invariance is broken by sudden bursts of elevated scale-dependent instantaneous dimensions $D(t, f)$, temporally localized differently at different frequencies, with also markedly different behavior of the local persistence properties. We clearly observe a transition towards a two-lobe attractor at low frequencies matching the expected phase-space geometry of the full attractor as well as the frequency break observed in the spectral properties. Thus, we have demonstrated that the geometrical and topological properties of invariant objects (i.e., attractors) are in fact both scale and time-dependent, being sensitive to the emergence of an intrinsic timescale solely determined by nonlinear interactions. For this reason, since the studied attractor adapts its geometric and statistical properties dynamically in time with respect to the intrinsic timescale, we suggest to call such an attractor a {\em chameleon attractor}. Furthermore, we also observed that the symmetric case has a very simple phase-space topology and can be associated with a noisy fixed point, while the non-symmetric case attractor displays scale-dependent features with a heterogeneous spatial distribution. 

Our results demonstrate that we cannot appropriately describe such attractors with full/averaged properties, and that we need refined analysis tools to detect their heterogeneity and the state-dependent properties of the system. Hence, it is apparent that the analysis of multiscale systems requires considering concepts allowing us to explore local and instantaneous properties of the system \cite{Faranda17,Alberti21}. Our analysis shows that the highly heterogeneous {\em chameleon attractors} discussed here could be common in high-dimensional dynamical systems as those encountered in climate sciences. We are confident that follow-up studies will further demonstrate their existence in such systems by exploiting the framework applied in the present work.

\section*{Acknowledgements}
This work was funded through ANR EXPLOIT, grant agreement no. ANR-16-CE06-0006-01 and ANR TILT grant agreement no. ANR-20-CE30-0035. VL acknowledges the support received from the EPSRC project EP/T018178/1 and from the EU Horizon 2020 project TiPES (Grant no. 820970). RVD has received funding by the German Federal Ministry for Education and Research via the JPI Climate/JPI Oceans project ROADMAP (grant no. 01LP2002B).

 \bibliographystyle{elsarticle-num} 
 \bibliography{Albertietal.bib}

\end{document}